\documentclass[conference]{IEEEtran}
\IEEEoverridecommandlockouts
\usepackage{cite}
\usepackage{amsmath,amssymb,amsfonts}
\usepackage{algorithmic}
\usepackage{graphicx}
\usepackage{textcomp}
\usepackage{url}
\usepackage{booktabs}
\usepackage{multirow}
\usepackage{colortbl}
\usepackage{caption}
\usepackage[numbers,sort&compress]{natbib}
\usepackage{hyperref}
\usepackage{bbding}
\usepackage{soul}
\usepackage{bibunits}
\usepackage{bm}
\usepackage{array}
\usepackage[table]{xcolor}
\usepackage{pdflscape}
\usepackage{makecell}
\usepackage{framed,multirow}
\usepackage{threeparttable}
\usepackage{amssymb}
\usepackage{algorithm}
\usepackage{listings}
\usepackage{pythonhighlight}
\usepackage{float}
\usepackage{mathtools}
\usepackage{cuted}
\makeatletter
\newcommand\footnoteref[1]{\protected@xdef\@thefnmark{\ref{#1}}\@footnotemark}
\makeatother

\newcolumntype{P}[1]{>{\centering\arraybackslash}p{#1}}
\newlength\savewidth

\def\arrvline{\hfil\kern\arraycolsep\vline\kern-\arraycolsep\hfilneg}
\definecolor{mygray}{gray}{.9}
\definecolor{Highlight}{HTML}{39b54a}

\newcolumntype{x}[1]{>{\centering\arraybackslash}p{#1pt}}
\newcolumntype{z}[1]{>{\raggedright\arraybackslash}p{#1pt}}
\definecolor{citecolor}{HTML}{0071BC}
\definecolor{linkcolor}{HTML}{ED1C24}

\newcommand{\ourmodel}{{\fontfamily{ppl}\selectfont LN-Gen}}
\def\BibTeX{{\rm B\kern-.05em{\sc i\kern-.025em b}\kern-.08em
    T\kern-.1667em\lower.7ex\hbox{E}\kern-.125emX}}

\DeclareRobustCommand*{\IEEEauthorrefmark}[1]{%
    \raisebox{0pt}[0pt][0pt]{\textsuperscript{\footnotesize\ensuremath{#1}}}}

\begin{document}

\title{LN-Gen: Rectal Lymph Nodes Generation via Anatomical Features\\
\thanks{*Corresponding author. \textbf{wansh@ustc.edu.cn}

This work is supported by The University Synergy Innovation Program of Anhui Province (Grant No. GXXT-2022-056).}
}

\author{
\IEEEauthorblockN{
Weidong Guo\IEEEauthorrefmark{1,2}
Hantao Zhang\IEEEauthorrefmark{1,2},
Shouhong Wan\IEEEauthorrefmark{1,2}$^*$
Bingbing Zou\IEEEauthorrefmark{3,2,4}
Wanqin Wang\IEEEauthorrefmark{3,2,4}, and
Peiquan Jin\IEEEauthorrefmark{1,2}}

\IEEEauthorblockA{\IEEEauthorrefmark{1}School of Computer Science and Technology, University of Science and Technology of China, Hefei, China}
\IEEEauthorblockA{\IEEEauthorrefmark{2}Institute of Artificial Intelligence, Hefei Comprehensive National Science Center, Hefei, China}
\IEEEauthorblockA{\IEEEauthorrefmark{3}Department of General Surgery, The First Affiliated Hospital of Anhui Medical University, Hefei, China}
\IEEEauthorblockA{\IEEEauthorrefmark{4}Anhui Medical University, Hefei, China} \\

{\tt\small  wansh@ustc.edu.cn
}
}

\maketitle

\begin{abstract}
Accurate segmentation of rectal lymph nodes is crucial for the staging and treatment planning of rectal cancer. However, the complexity of the surrounding anatomical structures and the scarcity of annotated data pose significant challenges. This study introduces a novel lymph node synthesis technique aimed at generating diverse and realistic synthetic rectal lymph node samples to mitigate the reliance on manual annotation. Unlike direct diffusion methods, which often produce masks that are discontinuous and of suboptimal quality, our approach leverages an implicit SDF-based method for mask generation, ensuring the production of continuous, stable, and morphologically diverse masks. Experimental results demonstrate that our synthetic data significantly improves segmentation performance. Our work highlights the potential of diffusion model for accurately synthesizing structurally complex lesions, such as lymph nodes in rectal cancer, alleviating the challenge of limited annotated data in this field and aiding in advancements in rectal cancer diagnosis and treatment. The code will be publicly available at \href{https://github.com/schmidtkk/LN-Gen}{https://github.com/schmidtkk/LN-Gen}.
\end{abstract}


\begin{figure*}[t]
\begin{center}
\includegraphics[width=\linewidth,scale=1.00]{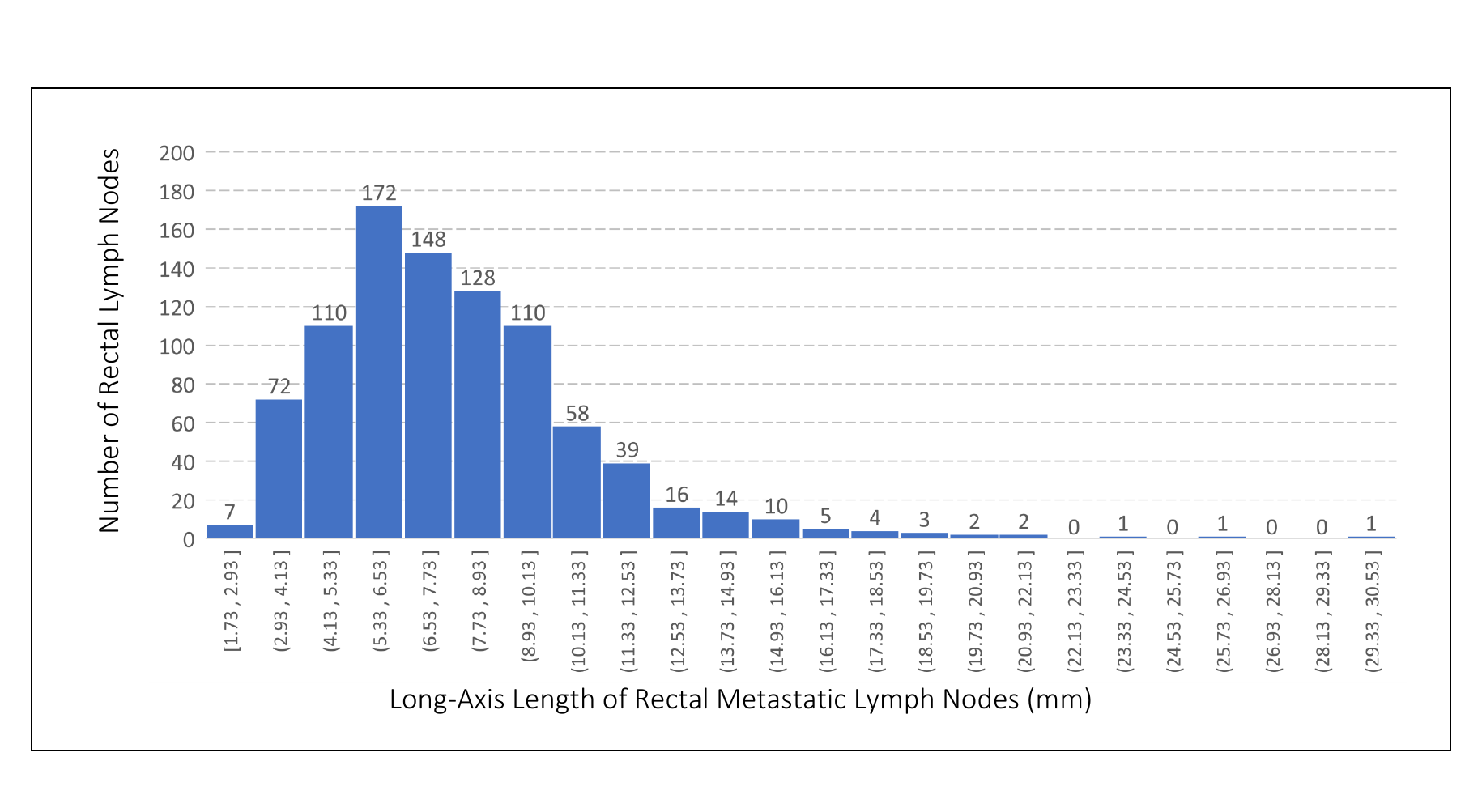}
\end{center}
\caption{\textbf{Distribution of the long-axis length of rectal lymph nodes in the training dataset.}
This figure illustrates the distribution of long-axis lengths for the 903 rectal lymph node samples within the training dataset. The long-axis lengths range from 1.7 mm to 30.0 mm, with the majority of samples measuring between 3 mm and 10 mm. Notably, samples with long-axis lengths exceeding 10 mm are scarce. This distribution reflects the considerable morphological and dimensional variability of rectal lymph nodes. Moreover, it underscores the pronounced imbalance within the dataset, which could potentially pose challenges for the effective training of segmentation models.}

\label{fig:distribution}
\end{figure*}
\section{Introduction}
The accurate segmentation of rectal lymph nodes is essential in assisting physicians with the staging of rectal cancer and the development of effective treatment plans~\cite{ganeshalingam2009nodal}. Precise lymph node segmentation can significantly enhance the diagnostic process, providing critical insights into the extent of cancer spread and facilitating personalized therapeutic strategies~\cite{rosenberg2010prognostic, rosenberg2008prognosis, tepper2001impact, koh2006nodal,zhang2023care}. Despite its importance, the task of segmenting rectal lymph nodes presents significant challenges due to the complexity and diversity of the surrounding anatomical structures.

Rectal lymph nodes are located in a region with a high density of various tissues and organs, each with distinct characteristics. This anatomical complexity makes manual annotation of lymph nodes a labor-intensive and time-consuming process, often requiring significant expertise and effort~\cite{zhang2023decoupling}. Furthermore, the reliance on manual annotations introduces variability and potential inaccuracies, which can affect the training and performance of machine learning models designed for segmentation tasks.

Another obstacle in developing robust segmentation models is the scarcity of annotated medical imaging data. High-quality, annotated datasets are crucial for training and validating machine learning algorithms. However, the collection and annotation of such data are constrained by the high costs and time requirements associated with manual labeling. This scarcity of data limits the potential for developing models that generalize well across diverse patient populations and imaging conditions.

Moreover, as shown in Figure~\ref{fig:distribution}, rectal lymph nodes exhibit considerable morphological diversity and size variability, further complicating the segmentation task~\cite{guo2024meply}. The variations in shape, size, and appearance of lymph nodes across different patients necessitate models that are highly adaptive and capable of handling this heterogeneity. Existing segmentation models often struggle to generalize across datasets with such intrinsic variability, leading to suboptimal performance in clinical settings.

To address these challenges, synthetic data generation techniques have emerged as a promising solution~\cite{yao2021label,zhang2024lefusion,lyu2022pseudo}. These techniques aim to augment limited datasets with artificially generated samples that mimic the properties of real anatomical structures. 
However, current lesion synthesis methods face several limitations. Endoscopic synthesis techniques, while effective in generating 2D video frames, may not fully encapsulate the complex 3D structure of lymph nodes~\cite{machavcek2023mask,dorjsembe2024polyp}. Additionally, current 3D tumor synthesis approaches often depend on pre-defined masks, which can limit the variety and realism of the synthesized lesions~\cite{jin2021free,hu2023label,chen2024towards}. These methods might also face challenges in producing lesions across a broad range of sizes, potentially resulting in datasets that may not provide the diversity required for training effective segmentation models.

In this work, we propose a novel lymph node synthesis technique that addresses these limitations by generating diverse and realistic synthetic rectal lymph node samples. 

\textbf{Anatomic Structure Generation via Implicit Diffusion and Explicit Adaptation.}
Given the significant morphological and size variations in rectal lymph nodes, coupled with a dataset heavily skewed in distribution~\cite{guo2024meply}, we propose generating samples of diverse shapes and sizes to achieve a more balanced data representation. To ensure the stability and authenticity of these synthesized structures, our generation network operates within an implicit space, capturing the intricate 3D anatomical structures of rectal lymph nodes. Additionally, we introduce an adapter in explicit space to reconstruct detailed surfaces, further enhancing the stability and accuracy of the generated structures.

\textbf{Rectal Lymph Node Synthesis Guided by Medical Priors.}
We incorporate anatomical information with medical priors to guide the synthesis process comprehensively. This approach enables our network to produce realistic rectal lymph nodes of varying shapes and sizes, accurately positioned within the appropriate background.

\textbf{Enhancing Segmentation Training with Synthetic Data.}
We construct a data pool comprising synthetic structures and background CT images with candidate locations. To introduce diversity, we randomly select elements from this pool to guide the synthesis process. The synthetic data is then combined with real data and used to train the segmentation model.

To validate the effectiveness of our approach, we conducted extensive experiments demonstrating that our synthesized data significantly enhances segmentation performance. The experimental results highlight the potential of our synthetic data generation method to improve the accuracy and robustness of rectal lymph node segmentation models.

In summary, our work tackles the key challenges in rectal lymph node segmentation, such as data scarcity, morphological diversity, and the limitations of existing synthesis methods. By employing implicit diffusion and explicit adaptation, guided by medical priors, we generate high-quality synthetic data. This data significantly improves the accuracy and robustness of segmentation models, ultimately leading to better diagnostic precision and treatment outcomes for rectal cancer patients.

\section{Related Work}

\subsection{Diffusion Models in Image Synthesis}

In recent years, diffusion models have achieved significant advancements in the synthesis of natural scenes~\cite{ho2022video,croitoru2023diffusion,yang2023diffusion}. These models have demonstrated their effectiveness in generating high-quality images, often surpassing traditional methods such as Generative Adversarial Networks (GANs) in tasks like image inpainting, super-resolution, and unconditional image generation.

One of the foundational works in this area was introduced by Ho et al., who developed the Denoising Diffusion Probabilistic Model (DDPM), which became a benchmark for generating realistic textures and structures from noise~\cite{ho2020ddpm}. Building on this, Song et al. proposed the Denoising Diffusion Implicit Model (DDIM), which offered higher denoising efficiency~\cite{song2020denoising}. Based on these foundational works, Dhariwal and Nichol demonstrated that diffusion models could achieve state-of-the-art results across various image synthesis benchmarks, establishing their superiority over GANs in terms of image fidelity and diversity~\cite{dhariwal2021diffusion}.

However, directly applying diffusion models to the synthesis of anatomical structures, such as rectal lymph node masks, presents challenges, including the potential generation of discontinuous or anatomically incorrect shapes, due to the inherent complexity and strict structural constraints of medical images. Specifically, when directly diffusing rectal lymph node masks, the model may fail to preserve the continuous and smooth boundaries required for accurate anatomical representation. To overcome these limitations, our work employs a latent space diffusion approach based on Signed Distance Functions (SDF), which preserves anatomical continuity and structural integrity throughout the diffusion process.

\subsection{Medical Image Synthesis}

Diffusion models have also emerged as a powerful tool in the domain of medical image synthesis. These models have demonstrated their ability to generate high-resolution medical images with improved fidelity~\cite{dhariwal2021diffusion}. For example, Iglesias et al. ~\cite{iglesias2024generation} developed a convolutional neural network to evaluate synthetic images generated by diffusion models, showing that these models can produce high-quality MRI and CT scans suitable for clinical applications.
Luo et al. ~\cite{luo2024measurement} proposed an uncertainty-guided diffusion model for medical image synthesis, which enhances the reliability and clinical relevance of the generated images. Moreover, Özbey et al. ~\cite{ozbey2023unsupervised} introduced an adversarial diffusion model specifically tailored for unsupervised medical image translation, addressing key challenges such as unpaired data translation.

\subsection{Lesion Synthesis in Medical Images}

Diffusion models have been widely used in the synthesis of lesions and other anatomical structures in medical images, playing a crucial role in augmenting training datasets and improving diagnostic algorithms~\cite{kazerouni2023diffusion}.
Dorjsembe et al. ~\cite{dorjsembe2023conditional} extended this work to 3D medical image synthesis, proposing a conditional diffusion model that produces high-quality 3D images for various clinical applications.

Despite these advances, synthesizing rectal lymph nodes poses unique challenges due to the complexity and variability of these anatomical structures. Traditional diffusion-based lesion synthesis techniques often struggle to capture these variations accurately. Our SDF-based diffusion approach addresses these challenges by generating anatomically diverse and accurate rectal lymph node structures, thereby enhancing the robustness of segmentation models trained on this synthetic data.

To address the challenges in synthesizing anatomically complex structures like rectal lymph nodes, we adopt an SDF-based latent space diffusion model. This approach ensures the preservation of anatomical integrity and generates a diverse set of high-quality synthetic images, which can be critical for improving the performance of downstream tasks such as segmentation and diagnosis.

\begin{figure*}[t]
\begin{center}
\includegraphics[width=\linewidth,scale=1.00]{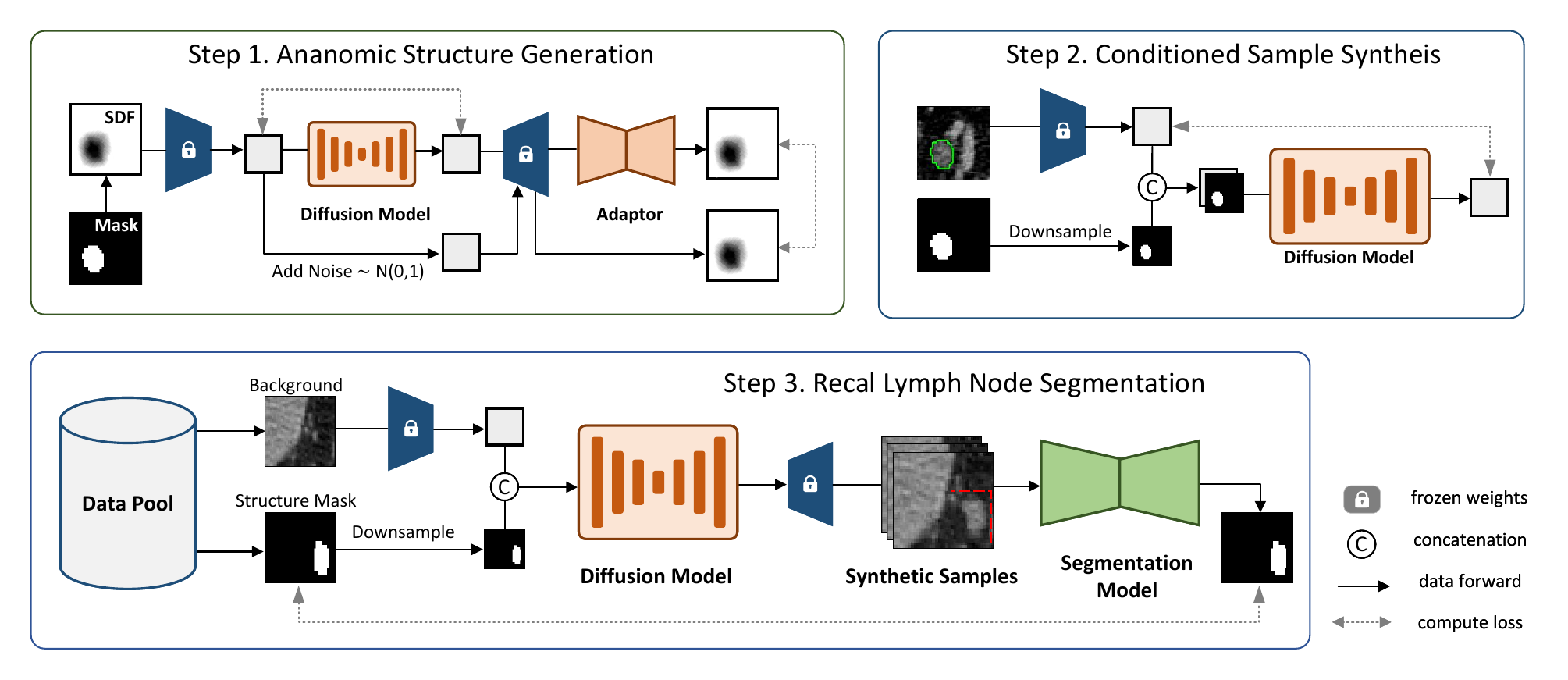}
\end{center}
\caption{\textbf{Overview of the proposed \ourmodel.}
This figure illustrates a three-step methodology for the synthesis and segmentation of rectal lymph node structures. In Step 1, the \textbf{Anatomic Structure Generation Network ($\phi$)} is trained on real anatomical data of rectal lymph nodes to generate synthetic structures. In Step 2, the \textbf{Conditioned Sample Synthesis Network ($\psi$)} is trained on real samples to synthesize rectal lymph node structures with realistic morphology and texture, guided by both morphological and positional information. In Step 3, the anatomical structures generated by $\phi$ are utilized to guide the generation of samples with diverse morphological structures, varying sizes, and realistic textures. These synthetic samples are then incorporated into the original training set to boost the segmentation model’s training.
}

\label{fig:framework}
\end{figure*}

\section{Methodology}

\subsection{Anatomical Structure Synthesis for Rectal Lymph Nodes}

To optimize computational efficiency and maintain the stability of the synthesis process, we utilize a Signed Distance Function (SDF) to model the intricate morphology of rectal lymph nodes. The adoption of an SDF-based approach allows for the precise capture of complex anatomical structures, which are subsequently encoded into a latent space. This latent representation forms the basis for the subsequent synthesis steps, enabling the generation of anatomically accurate models with high fidelity.

The synthesis process within this latent space is governed by a Denoising Diffusion Probabilistic Model (DDPM). The DDPM is particularly effective in addressing the inherent challenges associated with generating anatomically accurate and stable structures, as it iteratively refines the latent representations through a structured diffusion process. The forward diffusion process is formally defined as:

\begin{equation}
p(\mathbf{z}_t \mid \mathbf{z}_{t-1}) = \mathcal{N}(\mathbf{z}_t; \sqrt{1 - \beta_t} \mathbf{z}_{t-1}, \beta_t \mathbf{I}),
\end{equation}

where $\beta_t$ represents the variance schedule, controlling the degree of noise introduced at each timestep.

In the reverse diffusion process, the model incrementally denoises the latent representation, progressively reconstructing a refined anatomical structure. The noise prediction network, denoted as $\epsilon_\theta$, estimates the noise at each timestep $t$, and the corresponding diffusion loss is expressed as:

\begin{equation}
\mathcal{L}_{\text{diffusion}} = \mathbb{E}_{\mathbf{z}_0, \epsilon \sim \mathcal{N}(0,1), t} \left[ \left\| \epsilon - \epsilon_\theta (\mathbf{z}_t, \mathbf{z}_0, t) \right\|^2 \right],
\end{equation}

where $\mathbf{z}_0$ denotes the original latent code, and $\epsilon$ represents Gaussian noise.

To further enhance the quality of the synthesized anatomical structures and mitigate potential errors in the latent space, we introduce an anatomical adapter, denoted as $A_{\text{morph}}$. This adapter serves to refine the latent representation, thereby improving the reconstruction accuracy of detailed anatomical structures. Specifically, after encoding the input anatomical structure $\mathcal{M}$ into a latent code $\mathbf{z}$, Gaussian noise $\mathbf{G}$ is added, and the SDF decoder $D_{\text{SDF}}$ generates a noisy reconstruction $\hat{\mathcal{M}}$:

\begin{equation}
\hat{\mathcal{M}} = D_{\text{SDF}}(\mathbf{z} + \mathbf{G}),
\end{equation}

The anatomical adapter then processes $\hat{\mathcal{M}}$ to restore the original morphology $\mathcal{M}$, with the adapter loss formulated as:

\begin{equation}
\mathcal{L}_{\text{adapter}} = \left\| \mathcal{M} - A_{\text{morph}}(\hat{\mathcal{M}}) \right\|.
\end{equation}

The overall objective for training the rectal lymph node anatomical structure synthesis model is defined as a combination of the diffusion loss and the adapter loss:

\begin{equation}
\mathcal{L}_{\text{total}} = \mathcal{L}_{\text{diffusion}} + \lambda \mathcal{L}_{\text{adapter}},
\end{equation}

where $\lambda$ is a weighting factor that balances the contributions of these two components, ensuring both accurate latent space diffusion and high-fidelity morphological reconstruction. This methodology ultimately facilitates the synthesis of realistic and anatomically precise rectal lymph node structures, which are crucial for downstream medical imaging applications.

\subsection{Conditioned Synthesis of Rectal Lymph Nodes}

To generate anatomically realistic rectal lymph nodes that adhere to specific morphological constraints, we employ a conditional latent diffusion model. This model is designed to leverage control parameters that guide the synthesis process, thereby ensuring that the generated lymph nodes conform to desired anatomical characteristics.

The conditional latent diffusion model generates rectal lymph nodes by conditioning on an anatomical structure mask $\mathbf{m}$. During the reverse diffusion process, the latent representation is denoised while incorporating the control parameter $\mathbf{m}$. The conditional diffusion loss is defined as:

\begin{equation}
\mathcal{L}_{\text{cond-diff}} = \mathbb{E}_{\mathbf{z}_0, \epsilon \sim \mathcal{N}(0,1), t, \mathbf{m}} \left[ \left\| \epsilon - \epsilon_\theta (\mathbf{z}_t, \mathbf{z}_0, \mathbf{m}, t) \right\|^2 \right],
\end{equation}

where $\epsilon_\theta$ is the noise prediction network conditioned on $\mathbf{m}$, and $\mathbf{z}_0$ represents the original latent code. This conditioned approach ensures that the generated structures not only maintain anatomical accuracy but also align with the specific morphological requirements dictated by the clinical context.

\subsection{Incorporation of Medical Priors for Enhanced Anatomical Realism}

\textbf{Heuristic Selection of CT Background and Generation Regions.}  
To ensure that the synthesized rectal lymph nodes are both anatomically plausible and contextually appropriate, we integrate medical priors into the synthesis process. These priors inform the heuristic selection of suitable CT background regions by considering the typical anatomical locations and surrounding structures associated with rectal lymph nodes. By leveraging this domain-specific knowledge, we ensure that the generated lymph nodes are accurately positioned within their anatomical context, facilitating seamless integration with surrounding tissues. This method preserves anatomical accuracy and coherence within the synthetic images, thereby enhancing the reliability of the generated data for subsequent tasks such as segmentation.

\textbf{Scaling to Increase Anatomical Diversity.}  
To further improve the anatomical realism and diversity of the synthesized rectal lymph nodes, we implement a scaling mechanism guided by medical priors. These priors offer insights into the natural variability of lymph node size and morphology, allowing for adjustments to the dimensions of the anatomical structures during synthesis. By incorporating this variability, the scaling mechanism produces lymph nodes of varying sizes, more accurately reflecting the diversity observed in real clinical data. This diversity is essential for constructing a robust dataset that captures the full range of rectal lymph node morphology, thereby enhancing the effectiveness and generalizability of segmentation models trained on this synthetic data.

\section{Experiments and Results}
In this section, we conduct relevant experiments to evaluate our proposed \ourmodel. The experimental results are as follows.

\begin{figure*}[t]
\begin{center}
\includegraphics[width=\linewidth,scale=1.00]{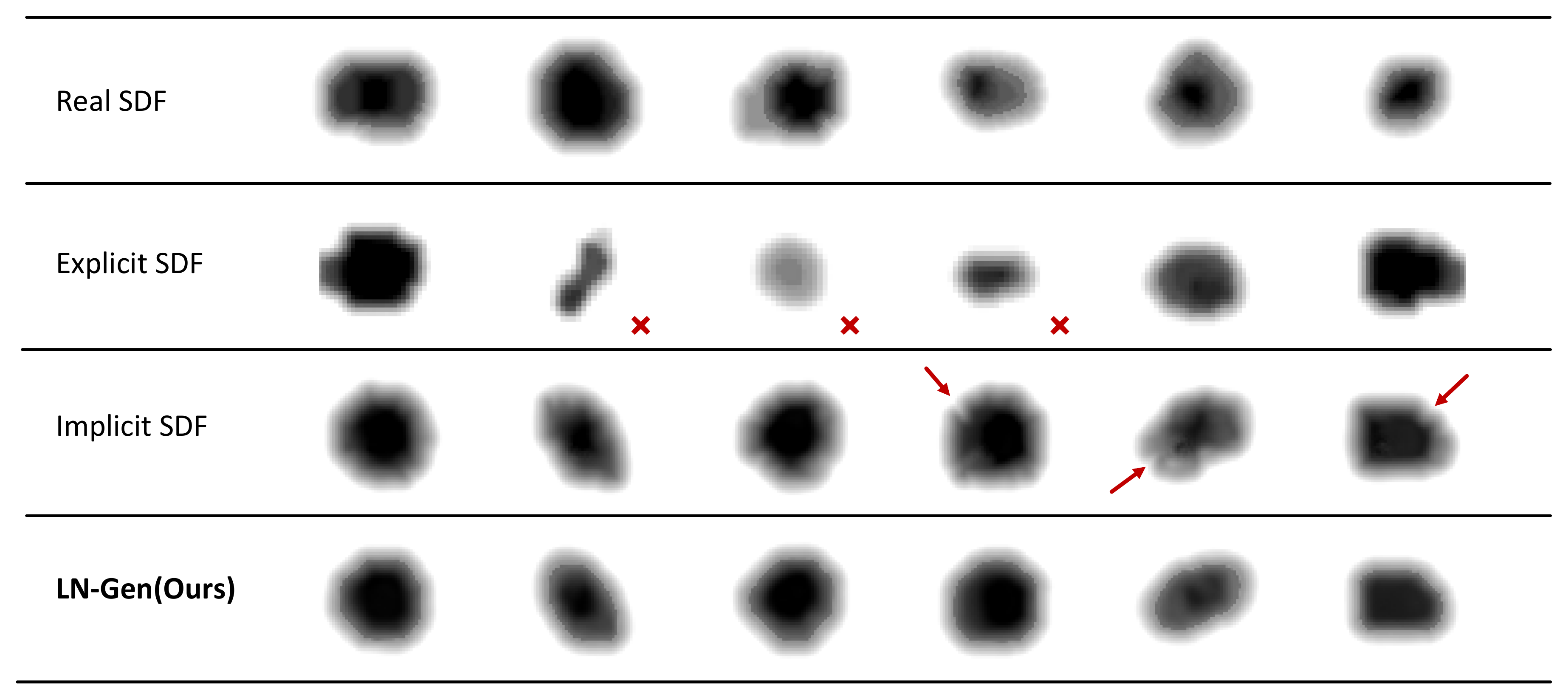}
\end{center}
   \caption{
    \textbf{The synthetic anatomical structures in SDF presentation.}
    We present both real and synthetic rectal lymph nodes in SDF representation. \ourmodel\ effectively generates anatomical structures with extensive diversity, remarkable stability, and high quality, ensuring the authenticity of the structures while accurately capturing the detailed surface information of the lymph nodes. Incorrectly generated structures are marked with a red X, and structures with minor flaws are indicated by red arrows.}

\label{fig:sdf-visualization}
\end{figure*}

\subsection{Dataset}
\textbf{Rectal lymph nodes dataset.} This dataset extends the Meply~\cite{guo2024meply} dataset by incorporating additional lymph node annotations to improve its comprehensiveness. The dataset comprises 120 contrast-enhanced CT scans, annotated with a total of 1,356 rectal lymph nodes. For model development, 80 scans with 903 rectal lymph nodes were allocated to the training set, 20 scans with 235 rectal lymph nodes to the validation set, and the remaining 20 scans with 218 rectal lymph nodes were reserved for testing.

You're correct that the phrase "improving the relevance of the training data" might not accurately convey the intended meaning. Here's a revised version:

\subsection{Implementation Details}

All experiments were conducted using PyTorch on a server equipped with an NVIDIA RTX 3090 GPU. For the synthesis of rectal lymph node structures, the segmentation mask was processed using a Truncated Signed Distance Function (Truncated-SDF), with values constrained between -0.2 and 0.2. This truncation focuses on the most relevant regions of the lymph nodes, reducing noise from less significant areas. A Variational Autoencoder (VAE) pre-trained on non-medical SDF data~\cite{li2023diffusion} was employed to encode the SDF into a latent space, capturing the essential features of the lymph node shapes. This latent representation was then used for diffusion within the latent space, followed by decoding to reconstruct the synthesized lymph node structures.

In synthesizing rectal lymph node samples, the Hounsfield Unit (HU) values of the CT images were truncated to a range of -175 to 250, effectively normalizing the CT data by emphasizing soft tissues while minimizing the influence of surrounding bone structures. To address the issue of morphological distribution imbalance among the synthesized samples, a random scaling strategy was implemented. This strategy not only varies the size of the samples but also introduces controlled randomness, enhancing the model’s ability to generate samples with diverse morphological characteristics.

During the generation of training data for segmentation tasks, synthesized structures were randomly selected from the pool of generated samples. A heuristic method was employed to determine appropriate locations within the CT images, considering anatomical context and spatial relationships. This step ensures that the synthesized lymph node samples are realistically positioned, thereby enhancing the anatomical plausibility of the training data. To maintain morphological diversity among the generated samples, the structures were scaled to achieve a uniform distribution of long-axis diameters, ranging from 1.7 mm to 30 mm. This scaling ensures that the synthetic dataset covers a wide range of lymph node sizes, which is critical for training segmentation models capable of handling the natural variability found in clinical data.

\subsection{Evaluation on Segmentation Task}
\begin{figure*}[t]
\begin{center}
\includegraphics[width=\linewidth,scale=1.00]{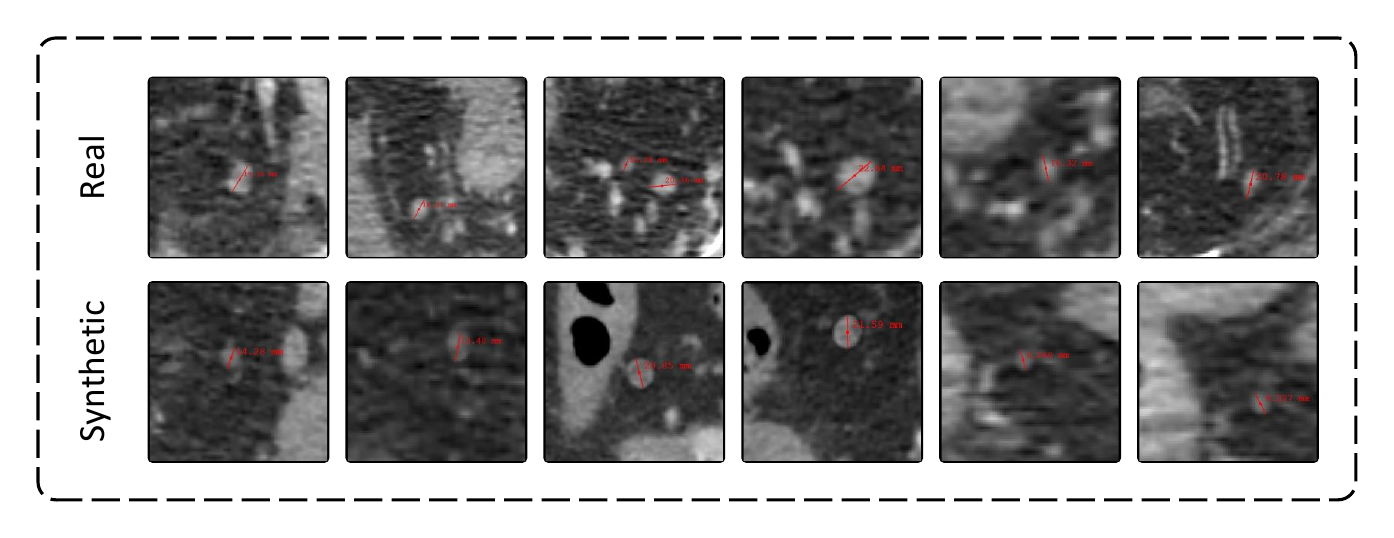}
\end{center}
\caption{\textbf{Visualization of real and synthetic rectal lymph nodes.} We present real and synthetic rectal metastatic lymph nodes, demonstrating that our approach can synthesize lymph nodes of varying sizes and morphologies with high quality and authenticity. The red lines indicate the long-axis length of the rectal lymph nodes.}
\label{fig:sample-visualization}
\end{figure*}

To rigorously assess the efficacy of our synthetic generation approach, we conducted comprehensive experiments on a downstream segmentation task utilizing three distinct training datasets: a dataset consisting solely of real lymph nodes, a mixed dataset combining real lymph nodes with DiffTumor-generated data, and a mixed dataset incorporating real lymph nodes with our LN-Gen synthetic data. All models were evaluated on a standardized test set to ensure consistent and equitable comparison.

As presented in Table~\ref{tab:compare_sota}, the quantitative results demonstrate that the integration of LN-Gen synthetic data into the training process significantly enhances segmentation performance across all evaluated networks, including U-Net~\cite{ronneberger2015u}, nnU-Net~\cite{isensee2021nnu}, and SwinUNETR~\cite{hatamizadeh2021swin}. Models trained with LN-Gen data consistently outperformed those trained exclusively on real lymph node data or on data augmented with DiffTumor-generated samples. Notably, while the addition of DiffTumor data did not consistently yield improvements in segmentation performance, the incorporation of LN-Gen data led to more robust and reliable enhancements across various architectures.

These findings underscore the critical importance and effectiveness of anatomically guided generation techniques in the synthesis of rectal lymph nodes. Our method is particularly well-suited for this application, given the inherent morphological diversity and significant size variation of rectal lymph nodes. The anatomical guidance embedded within our approach ensures that the synthesized structures are both realistic and diverse, effectively capturing the complex morphological characteristics of real rectal lymph nodes.

\begin{table*}[t]
    \centering
    \normalsize
    \begin{tabular}{p{0.15\linewidth}p{0.25\linewidth}P{0.2\linewidth}P{0.2\linewidth}} \\
    \toprule
    Network & Method & Sensitivity(\%) & Dice(\%)\\
    \midrule
    \multirow{3}{*}{\makecell[l]{U-Net~\cite{ronneberger2015u}}} & real lymph nodes & 48.07 & 50.49\\
    & DiffTumor~\cite{chen2024towards} &49.23 & 52.38\\
    & \ourmodel & \cellcolor{gray!10}\textbf{53.01} & \cellcolor{gray!10}\textbf{56.14}\\
    \midrule
    \multirow{3}{*}{\makecell[l]{nnU-Net~\cite{isensee2021nnu}~}} & real lymph nodes &50.42 & 54.24\\
    & DiffTumor~\cite{chen2024towards} &47.26 & 52.18\\
    & \ourmodel & \cellcolor{gray!10}\textbf{58.95} & \cellcolor{gray!10}\textbf{55.95}\\
    \midrule
    \multirow{3}{*}{\makecell[l]{SwinUNETR~\cite{hatamizadeh2021swin}}} & real lymph nodes & 51.84 & 55.47\\
    & DiffTumor~\cite{chen2024towards} & 43.48 & 50.81\\
    & \ourmodel & \cellcolor{gray!10}\textbf{55.42} & \cellcolor{gray!10}\textbf{56.48}\\
    \bottomrule
    \end{tabular}
    \vspace{5px}
    
    \caption{
    \textbf{The enhancement on segmentation performance: } 
    A comparison of segmentation performance using the real training set versus the real/synthetic mixed training set. The bold values indicate the best segmentation performance for each model under the same training data. Our method achieved the best performance across different models.
    }
    \label{tab:compare_sota}

\end{table*}

\subsection{Ablation Study}
We performed an ablation study to assess the impact of various generation strategies on the quality of synthesized rectal lymph node structures. To quantify the alignment between the distribution of synthetic samples and that of real data, we employed the Improved Precision and Recall metrics~\cite{kynkaanniemi2019improved}, referred to as IP and IR, respectively.

As presented in Table~\ref{tab:ablation}, the explicit modeling approach yielded synthetic structures of lower quality, characterized by coarse granularity and structural instability. Although implicit modeling produced more stable structures, it was limited by inadequately defined edges and a lack of fine detail. In contrast, the adapter-based approach markedly enhanced the synthesis process, generating high-resolution lymph node structures with well-defined edges and consistent morphological stability.
\begin{table}[t]
    \centering
    \normalsize
    \begin{tabular}{>{\centering\arraybackslash}m{0.15\linewidth} >{\centering\arraybackslash}m{0.15\linewidth}|>{\centering\arraybackslash}m{0.25\linewidth} >{\centering\arraybackslash}m{0.25\linewidth}} \\
    \toprule
    Implicit & Adapter & IP(\%) & IR(\%)\\
    \midrule
    & & 69.63 & 61.68\\
    \checkmark &  & 82.50 & 73.09 \\
    \checkmark & \checkmark & \cellcolor{gray!10}\textbf{86.38} & \cellcolor{gray!10}\textbf{76.52} \\
    \bottomrule
    \end{tabular}
    \vspace{5px}
\caption{
\textbf{Ablation Study: Diversity in Lymph Node Structure Synthesis.} The “implicit" column indicates the use of the implicit SDF method, while the “adapter" column signifies the introduction of an anatomical adaptor. Our approach, which incorporates the implicit SDF method and an anatomical adaptor, generates rectal lymph nodes that are more realistic and diverse, closely resembling the true distribution. 
}
    \label{tab:ablation}
\end{table}

\section{Conclusion}
This study proposes a methodology for synthesizing rectal lymph node structures using implicit Signed Distance Functions (SDF) augmented with an adapter to improve structural quality and realism. By addressing some of the limitations in existing synthesis techniques, our approach aims to generate more detailed and anatomically accurate representations of rectal lymph nodes, which are important for medical imaging applications.

Additionally, we developed an anatomically guided synthesis technique that incorporates medical priors to ensure that the generated samples maintain fidelity while reflecting the natural diversity in morphology and size characteristic of rectal lymph nodes. This approach seeks to better capture inherent variations, providing a more comprehensive dataset for further analysis.

The effectiveness of our method was evaluated through downstream segmentation tasks, where the synthetic lymph node samples showed distributions that closely aligned with real anatomical data. This suggests that our approach may help in replicating the morphological diversity needed for training reliable segmentation models, potentially improving their performance in clinical applications.

In summary, our methodology offers a potential improvement in the quality and usefulness of synthetic data for medical imaging, which could contribute to the development of more accurate and robust segmentation models, ultimately supporting better clinical outcomes.

\section*{Acknowledgment}
This work is supported by The University Synergy Innovation Program of Anhui Province (Grant No. GXXT-2022-056).

\newpage
\bibliographystyle{IEEEtran}
\small\bibliography{reference}

\end{document}